\documentclass{appolb}
\usepackage{graphicx}
\usepackage{slashed,amsmath,amssymb,wrapfig}
\newcommand{\nn}{\nonumber}
\newcommand{\beq} {\begin{equation}}
\newcommand{\eeq} {\end{equation}}
\newcommand{\beqa} {\begin{eqnarray}}
\newcommand{\eeqa} {\end{eqnarray}}

\newcommand{\as}{{\alpha_s}}

\newcommand{\la}{\Lambda}

\newcommand{\ieps}{i\varepsilon}

\newcommand{\order}[1]{${\cal O}\left(#1 \right)$}

\newcommand{\eq}[1]{(\ref{#1})}

\newcommand{\halft}{{\textstyle \frac{1}{2}}}

\newcommand{\ket}[1]{\left\vert{#1}\right\rangle}
\newcommand{\bra}[1]{\langle{#1}\vert}

\newcommand{\com}[2]{\left[{#1},{#2}\right]}
\newcommand{\acom}[2]{\left\{{#1},{#2}\right\}}
\newcommand{\tr}{\mathrm{Tr}\,}

\newcommand{\bs}[1]{\boldsymbol{#1}}

\newcommand{\xv}{{\bs{x}}} 
\newcommand{\yv}{{\bs{y}}}
\newcommand{\pv}{{\bs{p}}}

\newcommand{\Pv}{{\bs{P}}}
\newcommand{\gv}{\bs{\gamma}}
\newcommand{\nv}{\bs{\nabla}}

\newcommand{\xbj}{{x_{Bj}}}

\begin{document}
\title{Relativistic bound states at Born level%
\thanks{Based on work with D. D. Dietrich and M. J\"arvinen \cite{Dietrich:2012iy}. Presented at Light Cone 2012, Krakow, Poland.}%
}
\author{Paul Hoyer
\address{Department of Physics and Helsinki Institute of Physics\\ POB 64, FIN-00014 University of Helsinki, Finland}
}
\maketitle
\begin{abstract}

Theoretical and phenomenological studies indicate that the QCD coupling $\as(Q^2)$ freezes in the infrared. Hadrons may then be described by a perturbative expansion around ``Born'' states bound only by a confining potential. A linear potential results from the QCD equations of motion when Gauss' law for $A^0$ is solved with $F_{\mu\nu}^a F^{\mu\nu}_a \neq 0$ as boundary condition. The \order{\as^0} Born states are Poincar\'e covariant and can serve as $\ket{in}$ and $\bra{out}$ states of scattering amplitudes. Their Dirac-type wave functions include $f\bar f$ creation/annihilation effects giving sea-like partons at low $\xbj$.
\end{abstract}

\section{Bound states at \order{\alpha_s^0}}\label{sectone}

Hadrons are highly relativistic bound states. The mass difference between excited states is of the same order as light hadron masses, which in turn are much larger than the $u,d,s$ (current) quark masses. Parton distributions reveal the relativistic motion of quarks in the nucleon, and the presence of a non-vanishing sea quark distribution even at low scales \cite{Beringer:1900zz}. 

Relativistic dynamics and color confinement are often thought to imply that the QCD coupling $\as(Q^2)$ is large at small momentum scales $Q$. Hadrons nevertheless have features which seem difficult to reconcile with a strongly coupled theory. To name a few:
\begin{itemize}
\item Hadron spectra reflect their valence quark ($q\bar q$ and $qqq$) degrees of freedom. There is no firm evidence for exotic, glueball or hybrid states. The sea quarks do not manifest themselves in the excitation spectrum. 

\item The OZI rule \cite{Okubo:1963fa}. {\em E.g.}, the $\phi(1020)$ decays predominantly to $K\bar K$, even though this final state is barely allowed kinematically. In a strong coupling scenario one would expect little suppression of $s\bar s \leftrightarrow u\bar u, d\bar d$ transitions at the mass scale of the strange quark, $m_s \sim 100$ MeV. 

\item Perturbation theory explains many features of hadron production down to low momentum scales \cite{Dokshitzer:2010zza}.
\end{itemize}

Considerations like the above motivate studying the possibility that $\as$ is of moderate size even in the confinement domain. This is not as heretical as it may sound. Several theoretical and phenomenological studies \cite{Brodsky:2002nb} concur that the strong coupling freezes at a moderate value. The quark model gives a semi-quantitative understanding of hadrons using the perturbative QCD potential added to a spin-independent linear potential. Features like the $\Sigma-\Lambda$ mass splitting are then explained by single gluon exchange \cite{De Rujula:1975ge}.

How could the confining interaction be self-consistently described theoretically, and combined with perturbative QCD? One possibility is to impose a non-vanishing boundary condition on $F_{\mu\nu}^aF^{\mu\nu}_a$ in the solution of Gauss' law \cite{Hoyer:2009ep}. This can be illustrated in QED. Taking the diagonal matrix element of  $-\nv^2 A^0(\xv)= e\psi^\dag\psi(\xv)$, for a state where an electron is at $\xv_1$ and a positron at $\xv_2$, gives $4\pi A^0(\xv;\xv_1,\xv_2)=e/|\xv-\xv_1|-e/|\xv-\xv_2|$. The standard Coulomb potential is then $\halft[eA^0(\xv_1)-eA^0(\xv_2)]=-\alpha/|\xv_1-\xv_2|$.
If a non-vanishing field strength at spatial infinity is imposed,
\beq\label{Fboundcond}
\lim_{|\xv|\to\infty}F_{\mu\nu}^aF^{\mu\nu}_a(\xv)= -2\Lambda^4
\eeq
the solution of Gauss' law includes a homogeneous term,
\beq
A^0(\xv;\xv_1,\xv_2)= \Lambda^2\, \hat{\bs{\ell}}\cdot \xv+e/4\pi|\xv-\xv_1|-e/4\pi|\xv-\xv_2|
\eeq
where the unit vector $\hat{\bs{\ell}}(\xv_1,\xv_2)$ may depend on the positions of the electron and the positron but not on $\xv$. Stationarity of the action wrt. variations in $\hat{\bs{\ell}}$ sets $\hat{\bs{\ell}} \parallel \xv_1-\xv_2$. Thus arises an instantaneous confining interaction in the Hamiltonian \cite{Hoyer:2009ep},
\beq\label{hint}
H_{\Lambda}=-\frac{e\Lambda^2}{4}\int d\xv\, d\yv\, \psi^\dag\psi(t,\xv)|\xv-\yv|\psi^\dag\psi(t,\yv)
\eeq
Clearly we should set $\Lambda=0$ in QED. However, a similar analysis can be carried out for QCD, where a linear potential is called for by data, lattice calculations and the quark model. The boundary condition \eq{Fboundcond} provides a dimensionful parameter which is not present in the Lagrangian, and which can be determined to describe (although not explain) confinement. Gauge covariant bound states exist for color singlet $q\bar q$ mesons and $qqq$ baryons.

The linear potential in \eq{hint} is of \order{e\Lambda^2} and thus leading compared to the \order{e^2} perturbative potential. It was found to provide a Lorentz-covariant framework for bound states, giving energy eigenvalues with the correct dependence on the CM momentum \cite{Dietrich:2012iy,Hoyer:1986ei}. This is non-trivial for quantization at equal time, and indicates that the implementation of the non-vanishing boundary condition \eq{Fboundcond} preserves the Poincar\'e invariance. 

The road thus appears open to a perturbative expansion. The bound states formed by the linear interaction potential in \eq{hint} take the place of the free $\ket{in}$ and $\bra{out}$ states normally used in the scattering of pointlike particles. In effect, one perturbatively expands around ``Born terms'' which incorporate confinement but no perturbative interactions.

\section{Wave functions}\label{secttwo}

Relativistic dynamics necessarily involves pair creation and annihilation. This is manifest in the sea quark distribution of the proton, which persists down to low scales \cite{Beringer:1900zz}. Consequently, relativistic bound states have an infinite number of Fock components. This need not exclude an analytic description, as demonstrated by the states of an electron in a static Coulomb field. The bound state energies $E$ are given by the Dirac equation,
\beq\label{diraceq}
\big[-i\gamma^0\nv\cdot\gv+eA^0(\xv) +m\gamma^0)\big]\phi(\xv) = E \phi(\xv)
\eeq
which is obtained by summing all diagrams where the electron interacts with the external field.
As was recognized early on in the ``Klein paradox'' \cite{Klein:1929zz}, the Dirac wave function $\phi(\xv)$ includes $e^+e^-$ pair effects. Time-ordering of the electron interactions shows that scattering into negative energy states corresponds to pair creation and annihilation.

The time-independence of $A^0(\xv)$ in \eq{diraceq} implies that the bound state energies $E$ are unchanged if retarded (instead of Feynman) electron propagators are used in all diagrams\footnote{A time-independent external field does not transmit energy. Hence the  $p^0$ component of the electron momentum is preserved. If $p^0>-m$ the negative-energy pole of the electron propagator at $p^0=-\sqrt{\pv^2+m^2}$ is never probed. This makes the Green function $G(p^0,\pv)$ independent of the $\ieps$ prescription at that pole \cite{Hoyer:2009ep}.}. In retarded propagation only a single (positive or negative energy) electron is present at any time. The Dirac wave function $\phi(\xv)$ in \eq{diraceq} describes the electron with this boundary condition. In analogy to cross sections \cite{Baltz:2001dp}, retarded boundary conditions  give $inclusive$ rather than exclusive charge densities $\phi^\dag\phi(\xv)$.

Electron pairs contribute significantly to the Dirac charge density whenever the potential is strong (and the dynamics thus is relativistic). Their contribution generally makes the Dirac wave function unnormalizable \cite{plesset}. This holds for any potential which is a polynomial in $r$ or in $1/r$ -- except for $V(r) \propto 1/r$. Similarly in $D=1+1$ dimensions any potential that is a polynomial in $x$ or $1/x$ gives unnormalizable wave functions. The absence of the normalization condition $\int d^3\xv\,\phi^\dag\phi(\xv)=1$ makes the Dirac energy spectrum $continuous$, quite unlike the discrete Schr\"odinger spectrum\footnote{This important property is bypassed in most modern textbooks. See also Ref. \cite{titchmarsh}.}.

Fig. 1(a) shows the Dirac wave function in \eq{diraceq} for the QED$_2$ potential $eA^0(x)=\halft e^2 |x|$. Since $m/e=2.5$ the dynamics is nearly non-relativistic at low $|x|$, and close agreement with the corresponding solution of the Schr\"odinger equation is indeed found for $e|x| \lesssim 5$ if the Dirac solution is normalized to unity in this region. However, the Dirac wave function starts to oscillate when the potential reaches twice the electron mass, $e|x| \simeq 10$, indicative of contributions from $e^+e^-$ pairs. Since $\phi(x \to \infty)\sim \exp(ie^2x^2/4)$ the Dirac charge density is asymptotically constant.


\begin{figure}[htb]
\centerline{%
\includegraphics[width=12.5cm]{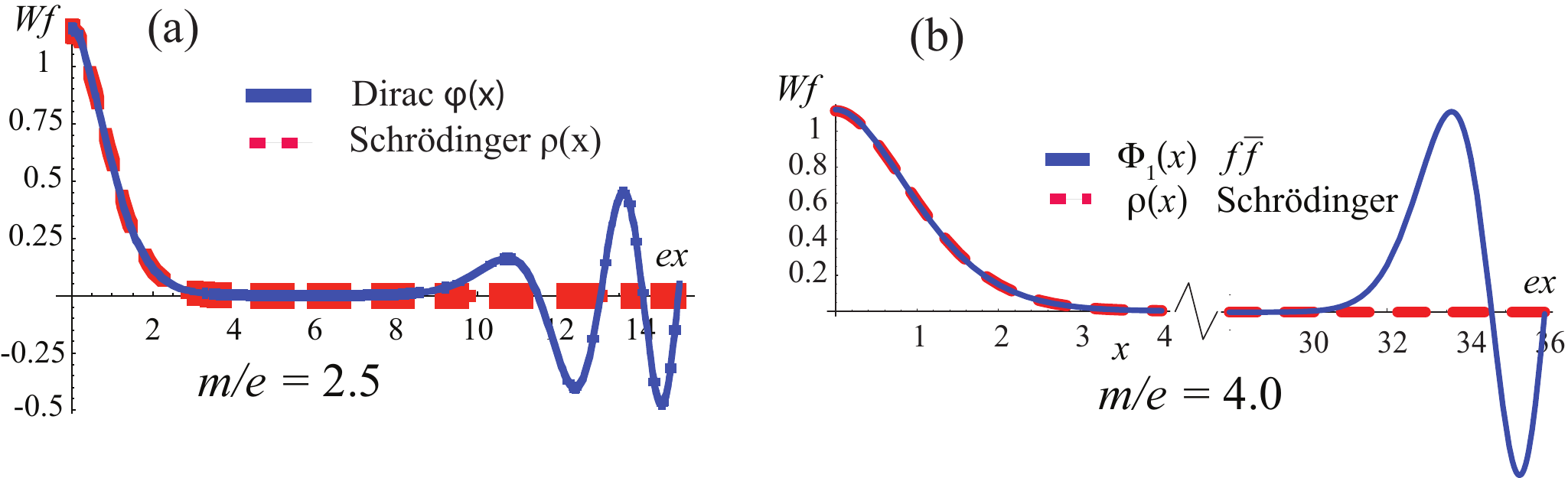}}
\caption{Wave functions in $D=1+1$ dimensions \cite{Dietrich:2012iy}. (a) Comparison of the upper component $\varphi(x)$ of the Dirac wave function \eq{diraceq} with the Schr\"odinger wave function $\rho(x)$ for $m/e=2.5$. (b) Comparison of one component of the $f\bar f$ wave function \eq{bse1} (for $m/e=4.0$) with the Schr\"odinger wave function (for $m/e=2.0$).}
\label{fig1}
\end{figure}

Retarded boundary conditions may plausibly be used with the instantaneous linear potential\footnote{But not with perturbative Coulomb photon/gluon exchange, which (for finite fermion masses) transmits energy as well as 3-momentum. See footnote 1.} \eq{hint}. Then the wave function $\Phi(\xv)$ of an $f\bar f$ bound state with CM momentum $\bs{P}$,
\beq\label{emustate}
\ket{P,t}=\int d\xv_{1}d\xv_{2}\,\bar\psi_{1}(t,\xv_{1})\exp\big[i\Pv\cdot (\xv_1+\xv_2)/2\big]\Phi(\xv_{1}-\xv_{2})\psi_{2}(t,\xv_{2})\ket{0}_{R}
\eeq
satisfies (for $m_1=m_2=m$)
\beq\label{bse1}
i\nv_x\cdot\acom{\gamma^0\gv}{\Phi(\xv)}-\halft \Pv\cdot\com{\gamma^0\gv}{\Phi(\xv)}+m\com{\gamma^0}{\Phi(\xv)} = \big[E-V(\xv)\big]\Phi(\xv)
\eeq
where $V(\xv) = \halft e\la^2 |\xv|$. The wave function $\Phi(\xv)$ is generally singular at $E=V(\xv)$. Requiring $\Phi(\xv)$ to be  regular at this point makes the energy spectrum discrete rather than continuous as in the Dirac case. The $f\bar f$ states in $D=1+1$ were found to transform correctly under boosts \cite{Dietrich:2012iy}, and the energy eigenvalues of \eq{bse1} satisfy $E=\sqrt{\Pv^2+M^2}$ \cite{Hoyer:1986ei}. Fig. 1(b) shows one component of the $2\times 2$ wave function $\Phi(x)$ for $m/e=4$, compared to the Schr\"odinger wave function with the reduced mass $m/e=2$. The comparison is qualitatively similar to the Dirac case in Fig. 1(a).
%
\begin{figure}[htb]
\centerline{%
\includegraphics[width=12.5cm]{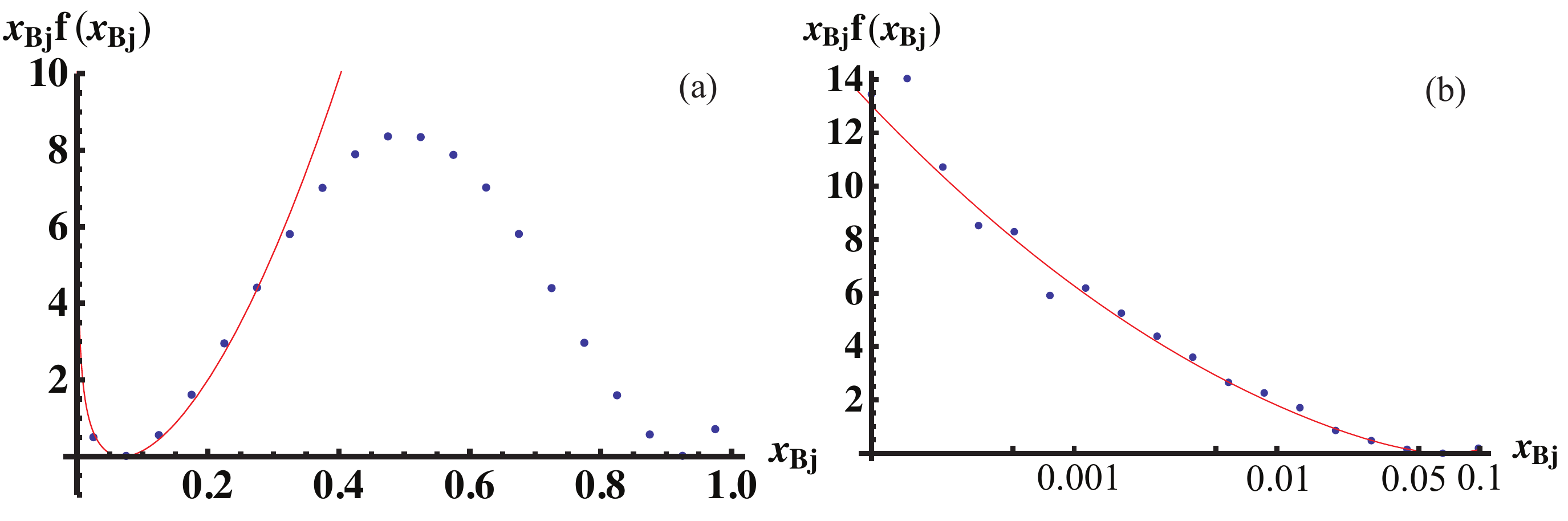}}
\caption{(a) $f\bar f$ ground state parton distribution in $D=1+1$ for $m/e=0.1$ (preliminary, \cite{Dietrich:2012iy}). (b) The same distribution on a logarithmic scale. The dots are numerical results and the curve shows an analytic approximation valid at low $\xbj$.}
\label{fig2}
\end{figure}

\vspace{-.5cm}
\section{Form factors and parton distributions}\label{sectthree}

The matrix element of the electromagnetic current $j^\mu(z)= \bar\psi(z)\gamma^\mu\psi(z)$ between $f\bar f$ bound states \eq{emustate} gives the form factor \cite{Dietrich:2012iy},
\beqa\label{formfac}
F^\mu_{AB}(z) &\equiv& \bra{B(P_b)}j^\mu(z)\ket{A(P_a)} \nn\\
&=& e^{i(P_b-P_a)\cdot z}\int d\xv\, e^{i(\Pv_b-\Pv_a)\cdot\xv/2}\,
\tr\big[\Phi_B^\dag(\xv)\gamma^\mu\gamma^0\Phi_A(\xv)\big]
\eeqa
Gauge invariance, $\partial_\mu F_{AB}^{\mu}(z) =0$, holds as a consequence of the bound state equation \eq{bse1} satisfied by the wave functions $\Phi_A, \Phi_B$.

The quark distribution of target state $A$ is obtained in the Bjorken limit where the photon virtuality and the mass of the final state $B$ tend to infinity. Since all states have zero width (before the perturbative corrections) an averaging procedure needs to be applied. The relative normalization of the wave functions $\Phi_B$ can be determined from duality between the contributions of bound states and free quarks to the imaginary parts of current propagators. Our preliminary result for the quark distribution of a relativistic ground state ($m/e=0.1$) in $D=1+1$ is shown in Fig.~2. The rise of the distribution at low $\xbj$ is attributed to $f\bar f$ pairs, indicating again the inclusive nature of the wave functions obtained with retarded boundary conditions. 

\vspace{.2cm}

\noindent {\bf Acknowledgements}

This presentation is based on a collaboration with D. D. Dietrich and M. J\"arvinen. I have benefitted from discussions with S. Brodsky. I am grateful to the organizers of Light Cone 2012 for their invitation, and for a travel grant from the Magnus Ehrnrooth Foundation.

\end{document}